\documentclass[aps,prb,showpacs,preprint]{revtex4}
\usepackage{graphicx}
\begin{document}
\def\mb{{\bf M}}
\def\hb{{\bf H}}

\title{Superradiation from Crystals of High-Spin Molecular Nanomagnets}

\author{V.K. Henner} \author{I.V. Kaganov}
\affiliation{Perm State University, 614600 Perm, Russia, and\\
 University of Louisville, Louisville, KY, 40292, U.S.A.}
\date{\today}

\begin{abstract}
Phenomenological theory of superradiation from crystals of
high-spin molecules is suggested. We show that radiation
friction can cause a superradiation pulse and investigate
the role of magnetic anisotropy, external magnetic field
and dipole-dipole interactions.
Depending on the contribution of all these factors at low temperature,
several regimes of magnetization of crystal sample are described.
Very fast switch of magnetization's direction for some sets of
parameters is predicted.
\end{abstract}
\maketitle
\section{Introduction}
Comparatively recently discovered \cite{l,fstz,ppbsc,hbtcb}
molecular nanomagnets of high-spin molecules are prospective
from the point of view of practical application of the
superradiation (SR) phenomena in electron spin
resonance (ESR) frequency region.
These are crystals of organic molecules which consist of about
$10$ atoms of transitional metals. They are bounded in the
molecule with indirect exchange interaction,
as a result, each molecule possesses spin $S\approx 10$.

One of the manifestations of SR phenomena
is the radiation intensity approximately proportional
to the number of radiators squared\cite{d}.
The properties of such coherent effects in optics
have been extensively studied.
Similar effects exist in the radiofrequency region
(SR in spin systems was observed in several experiments, see
Refs. \onlinecite{bbm,kpsy}),
but although these and optical phenomena are similar in some aspects,
the physics governing SR in spin systems is very different.
It is natural and rather helpful to use magnetic resonance
methods to describe SR in such systems\cite{bh1,bh2}.

Let us return to high-spin molecular nanomagnets.
The molecules interact mostly through dipole forces.
In macroscopic terms,
the motion of the magnetic
moment of a crystal composed of such molecules leads to the
radiation of electromagnetic waves, whose feedback influence
creates radiation friction force (Lorents force).
At low temperature, when the time of transverse relaxation
$T_2$ (most usually due to spin-spin interactions) is large,
relaxation by means of radiation friction can serve as
the main cause of SR. Pure electromagnetic radiation mechanism
(with $T_2^{-1}=0$) was proposed recently\cite{cg}
(see also discussion in Ref. \onlinecite{ylp}).
Both of the relaxation mechanisms mentioned above are
taken into account in the approach we elaborate in this work
for molecular nanomagnets.
Electromagnetic radiation is described in terms of the
retarded potential expansion; for the transverse relaxation
a regular magnetic resonance description is used.
Recently we applied similar description for nuclear and electron
SR in solids\cite{dkh}.

First, give some qualitative estimations.
The characteristic relaxation time due to radiation from a spin system
in constant external magnetic field $\hb_0$ is
$T_R\sim {c^3\over \gamma\omega_0^3 M}$,
where $M$ is a magnetic moment of the sample,
$\omega_0=\gamma H_0$, $\gamma$ is gyromagnetic ratio.
This time is obtained by the
expansion of the radiation field in powers of $1/c$
(see, for instance, Ref. \onlinecite{sk}).
The time $T_2$ due to dipole-dipole interactions
can be estimated (for simplicity assume paramagnetic with regular
structure) via the second moment of resonance line, $M_2$.
This time is about $1/\sqrt{M_2} \;$, where
$M_2\sim\gamma^2 n^2 \mu^2
\left( 1-L^2\left ({\mu H_0\over T}\right )-{2T\over\mu H_0}
L\left ({\mu H_0\over T}\right )\right )$ for large spin
(see Ref. \onlinecite{dkh}, where this formula was discussed).
Here $n$ is concentration of spins (number of
spins over number of lattice nodes in crystal), $\mu$ is
the magnetic moment of spin, $T$ is the temperature, $L(x)$ is
the Langevin function. If $T\to 0$, then $T_2\sim {H_0\over\gamma nT}$.
We discuss the situation when radiation friction is the main
relaxation mechanism, thus $T_R\ll T_2$.
Relaxation must be slow in comparison with spins rotation
about the field $\hb_0$, $\omega_0\gg {1\over T_R}$.
Also, the sample size should be smaller than the radiating wavelength,
${\omega_0V^{1/3}\over c}\ll 1 \;$
($V$ is the volume of the sample),
then the phase of emitted photons is the same throughout the sample and
the process of radiation becomes coherent.
In typical EPR experiments,
$\omega_0\sim 10^{12}{\rm s}^{-1}$ and $T\sim 0.1{\rm K}$,
and for linear size of a sample about $0.01{\rm cm}$
all these conditions may be satisfied.
The temperature can be higher for bigger $\omega_0$
and smaller sample size.
Thus, the need for SR situation when radiation damping
prevails over transverse relaxation is quite realistic for ESR.

Considering the high-spin molecule as a single paramagnetic particle,
a crystal sample is a super-paramagnet with strong magnetic
anisotropy\cite{ppbsc,nscg,gcs}. For example, the molecules
${\rm [Mn_{12}O_{12}(CH_3COO)_{16}(H_2O)_4]\cdot 2CH_3COOH\cdot 4H_2O}$
have spin $S=10$ and strong single-site anisotropy barrier about $70{\rm K}$.
This temperature corresponds to magnetic energy of such a spin
in a field of about $10^5{\rm G}$
(in this field $\omega_0\sim 10^{12}{\rm s}^{-1}$) that
leads to a shift of the resonant frequency due to $M_z$ variations.

\section{Equation of motion}
Equation of motion of the magnetic moment $\mb$ of the sample
of spherical
shape can be written as follows:
\begin{equation}
\label{e1}
\dot{\mb}=\gamma\mb\times\left (
\hb_0+\beta{M_z\over V}\hat{z}-{1\over 2c^2V^{1/3}}\lambda_{ij}\ddot{M}_j+
{2\over 3c^3}\stackrel{{\bf\ldots}}\mb
\right )+{\bf R}
\end{equation}
where $\hb_0$ is constant magnetic field, directed along the anisotropy axis
$Oz$, $\hat{z}$ is the unit vector along this axis, $\beta$ is the
anisotropy coefficient, $\lambda_{ij}$
is the tensor (which in general is determined by the shape of the sample),
${\bf R}$ is the term describing the transverse Bloch relaxation.
The second term in the brackets is
the anisotropy field (like in Landau --- Lifshits
equation\cite{ll8}), third and fourth terms appear as expansion
of the retarded potential by powers of $\omega_0V^{1/3}/c$.
The third term gives a small shift of the resonance frequency,
$\Delta\omega_0 \ll \omega_0$. It does not lead to any
noticeable physical effect and can be neglected.
The fourth term describes the electromagnetic friction.

Consider the case when the field $\hb_0$ and the anisotropy field are
large compared to all other fields in the crystal. Thus, the rotation of
$\mb$ around the effective field $\hb_0+\beta{M_z\over V}\hat{z}$ is the
fastest motion and all other effects are comparatively slow.
Then, the zero-th order approximation is
\begin{equation}\label{e11}
\dot{\mb}=\gamma\mb\times
\left (\hb_0+\beta{M_z\over V}\hat{z}\right ) .
\end{equation}
Using  (\ref{e11}) repeatedly, one can obtain
\begin{eqnarray}
\label{2a}
\ddot{\mb}=\gamma^2\left(
\left (\hb_0+\beta{M_z\over V}\hat{z}\right )
\left (\mb\cdot\left (\hb_0+\beta{M_z\over V}\hat{z}\right )\right )
-\mb\cdot\left (\hb_0+\beta{M_z\over V}\hat{z}\right )^2
\right ),
\\
\nonumber
\stackrel{{\bf\ldots}}\mb =-\gamma^3
\left (\hb_0+\beta{M_z\over V}\hat{z}\right )^2 \mb\times
\left (\hb_0+\beta{M_z\over V}\hat{z}\right ).
\end{eqnarray}
Substitute these expressions in (\ref{e1}) and then average equation
(\ref{e1}) over the fast frequency $H_0+\beta M_z/V$.
Finally we obtain:
\begin{eqnarray}
\label{e2}
\rho^2+n_z^2=1+{1\over K\xi}\left [{1\over (1+Kn_z)^2}-{1\over (1+Kn_z(0))^2}
\right ],\\
\label{e22}
\dot{n}_z=\xi (1-n_z^2)(1+Kn_z)^3+{1\over K}\left [
1+Kn_z-{(1+Kn_z)^3\over (1+Kn_z(0))^2}
\right ].
\end{eqnarray}
Here  $n_{x,y,z}=M_{x,y,z}/M(0)$ are normalized components of
the magnetic moment, $\rho^2=n_x^2+n_y^2\;,\;$
$K={\beta M(0)\over VH_0}$ is the dimensionless coefficient of
anisotropy, $\xi =T_2/T_R$ is the ratio of the
characteristic time of transverse relaxation and the characteristic time
of longitudinal relaxation due to radiation friction,
$T_R={3c^3\over 2\gamma\omega_0^3M(0)}$,
$n_z(0)=\cos\theta (0)$ where $\theta (0)$ is
the angle of the initial deviation of $\mb$ from the $Oz$ axis.
Note, that we measure the dimensionless "time" in the units of $T_2$.
We neglect
spin-lattice interaction here.

The intensity of dipole radiation\cite{ll2}
is  $I={2\over 3c^3}{\ddot{\mb}}^2$.
The biggest contribution in $\ddot{\mb}$ is due to rotation about
$\hb_0$, i.e. due to $\ddot{M}_x$ and $\ddot{M}_y$.
Using formulae (\ref{2a}) for these quantities, we obtain
\begin{equation}
\label{e3}
I={2M^2(0)\over 3c^3}\omega_0^4(1+Kn_z)^4\rho^2.
\end{equation}
It is seen that $I$ is proportional to the second power of the number of
molecules, $N$, in the sample which is $N$ times the incoherent radiation.
This is the manifestation of Dicke's superradiance for molecular nanomagnets.
Using the same parameters as in the Introduction, one can see that $I$ can be
as large as $10^{9}{\rm erg/s}$
for the concentration of molecules in crystal $n\approx
4\cdot 10^{20}{\rm cm}^{-3}$.
It is rather big quantity,
although this radiation splashes only during about $10^{-9}{\rm s}$
(because the amount of energy released during a flip of the
magnetic moment in the field $\hb_0$ is about 1~erg).
It is questionable whether such a short signal can be detected,
however, the flip of a total magnetic moment makes a direct
experimental observation of the phenomena possible.

\section{Regimes of motion}
\begin{figure}[ht]
\centerline{
\includegraphics*[scale=1.0]{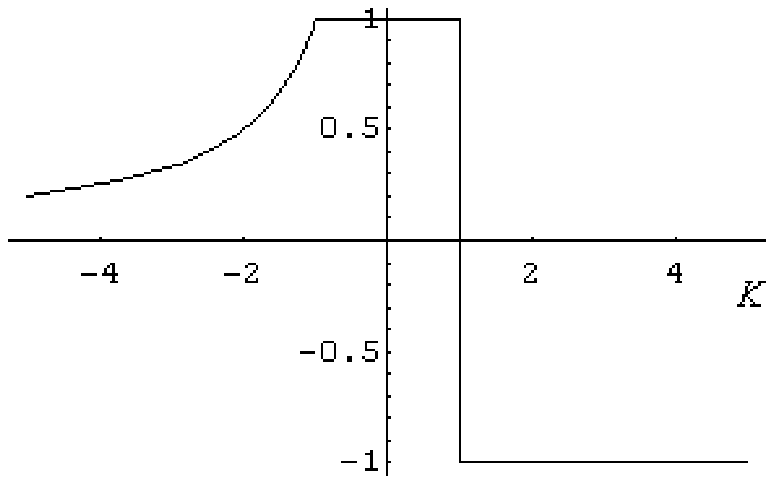}
}
\caption{The dependence of the final $z$-component of polarization on $K$
for $T_2^{-1}=0$. The value  $n_z(t=0)\approx -1$.}
\label{f1}
\end{figure}

Let us consider first the SR when transverse relaxation
can be completely neglected compared to radiation relaxation, i.e.
when $\xi\to\infty$.
This situation can be realized at low temperatures.
Fig.\ref{f1} shows how the
final (in the equilibrium state) $z$-component of polarization,
$n_z(t\to\infty)$, depends on the coefficient of anisotropy, $K$.
The evolution of $n_z(t)$ and $I(t)/I(0)$ for big but finite
$\xi$ (the value $\xi=100$ was chosen)
and several values of parameter $K$ is shown in Fig.\ref{f2}.
We want to attract attention to the case when the value
of $K$ is slightly less than $1$, in this situation the
magnetization switches its direction rapidly from the unequilibrium
state causing a very sharp and high SR pulse.
This can be obtained directly from equation (\ref{e22}) that
gives the evolution of the $z$-component
of the total magnetization.
Introduce the quantity $\epsilon (t) = \cos(\pi /2 - \theta (t))$
where $\theta (t)$ is the angle between vector ${\bf n}$ and the $z$ axis.
If initially vector ${\bf n}$ had only a small deviation from
positive or negative directions of the $Oz$ axis, then
$\epsilon (0)\equiv \epsilon_0 \ll 1$.
Substituting $n_z(t)=\pm 1\mp\epsilon (t)$ into (\ref{e22}),
for $K=1$ and $\xi\gg 1$, we obtain
\begin{eqnarray}\nonumber\epsilon (t)={\epsilon_0
\over (1-6\epsilon_0^3\xi t)^{1/3}}\end{eqnarray}
when ${\bf n}$ was directed (approximately) against $\hb_0$,
and
\begin{eqnarray}\nonumber
\epsilon (t)=\epsilon_0e^{-16\xi t}\end{eqnarray}
if ${\bf n}$ was directed (approximately) along $\hb_0$.
Thus, in unequilibrium case, the magnetic moment begins to deviate
very slowly from the negative direction of the $Oz$ axis and then,
at some moment, makes a very quick flip producing
a short powerful splash of SR.
Note, that the final value of
$\vert n_z\vert$ is smaller than $\vert{n_z(t=0)}\vert$
due to transverse relaxation.

The value of $K$ depends on the external field and, therefore,
can be tuned. The peak of SR exists in the region $0\leq K\leq 1$
(anisotropy of the type "easy axis") and
in the region $K<0$ ("easy plane"). For the values of $K\ll -1$,
the initial $z$-component of polarization generally is not
big, that makes the SR effect weaker. If $K>1$, no SR pulse appears.

\begin{figure}[ht]
\centerline{
\includegraphics*[scale=0.7,angle=-90]{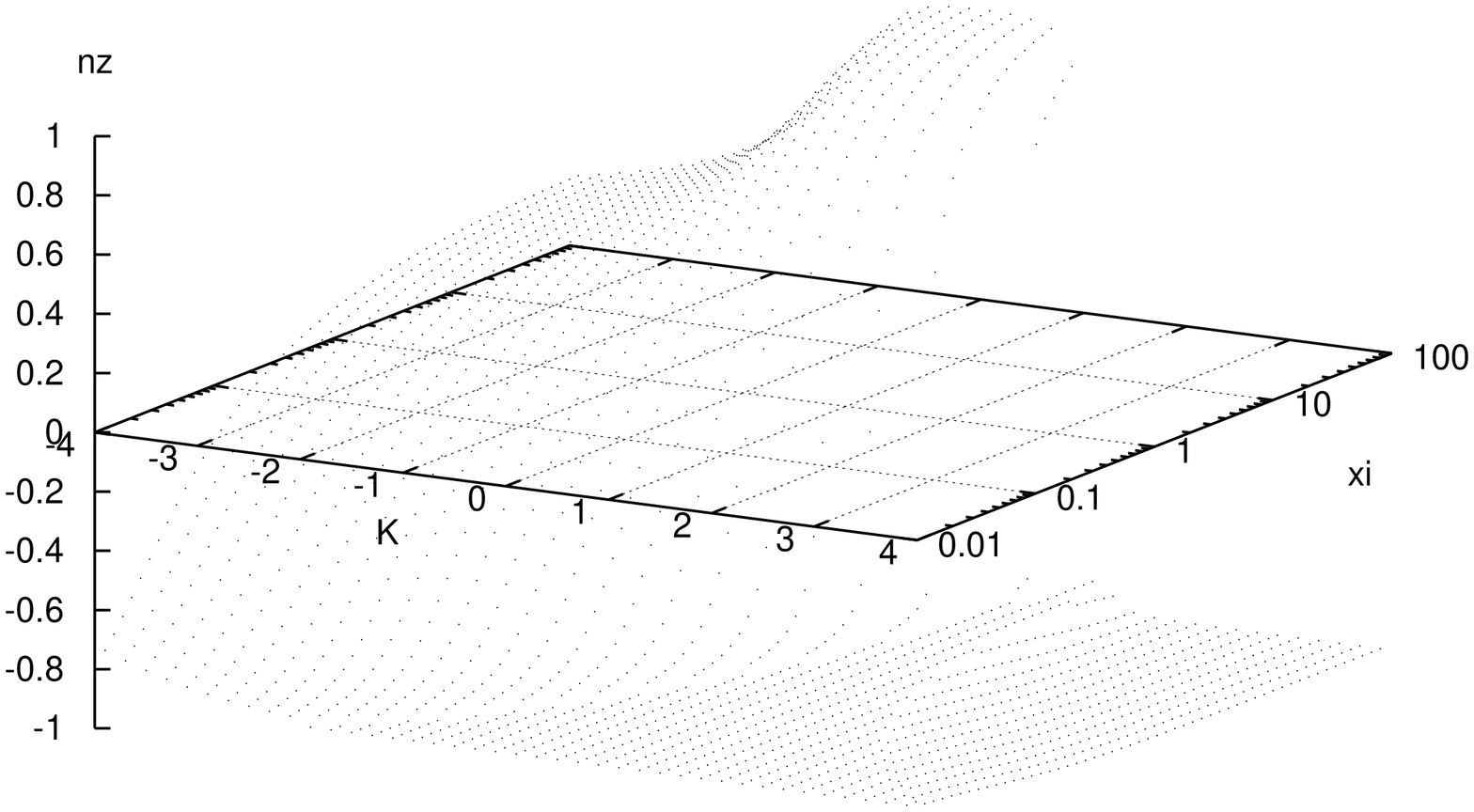}
}
\caption{The dependence of $n_z(t\to\infty ,\xi ,K)$
on $\xi$ and $K$ for $n_z(t=0)=-0.9$. }
\label{f3}
\end{figure}

When a relative contribution of the transverse relaxation increases,
the SR effect becomes smaller.
Fig.\ref{f3} is the 3-D plot of the final polarization
$n_z(t\to\infty)$
as function of $K$ and $\xi $
(for the initial polarization $n_z(t=0)=-0.9$).
Values of $n_z(t\to\infty)$
close to $-1$ mean that $\mb$ did not flip;
values close to $1$ indicate that
$\mb$ had changed its initial direction to the opposite.
It is seen that strong transverse relaxation (small values of $\xi$)
destroys the SR completely.

\section{Conclusion}
The usual way to obtain a SR pulse from spin system
is to place a sample in a tuned resonant coil.
A feedback from the coil creates a coherent relaxation from
the initially unequilibrium state.
However, for crystals of high-spin molecules a strong
magnetic anisotropy makes this mechanism not effective
because of the
significant variations in the resonant frequency\cite{ylp}.

In this paper we give a rigorous (phenomenologically based)
description for SR phenomena
in crystals of high-spin molecules. The main mechanism for the
SR is an interaction of spin magnetic moments with the cooperative
radiated electromagnetic field acting on an entire sample.
Contrary to this, the transverse relaxation changes phases
of spins rotation chaotically.
As a result, the radiation field can
switch the direction of $\mb$ and produce the SR pulse, while
the usual relaxation just dissipates the total magnetization.

\begin{figure}[ht]
\centerline{
\includegraphics*[scale=1.0]{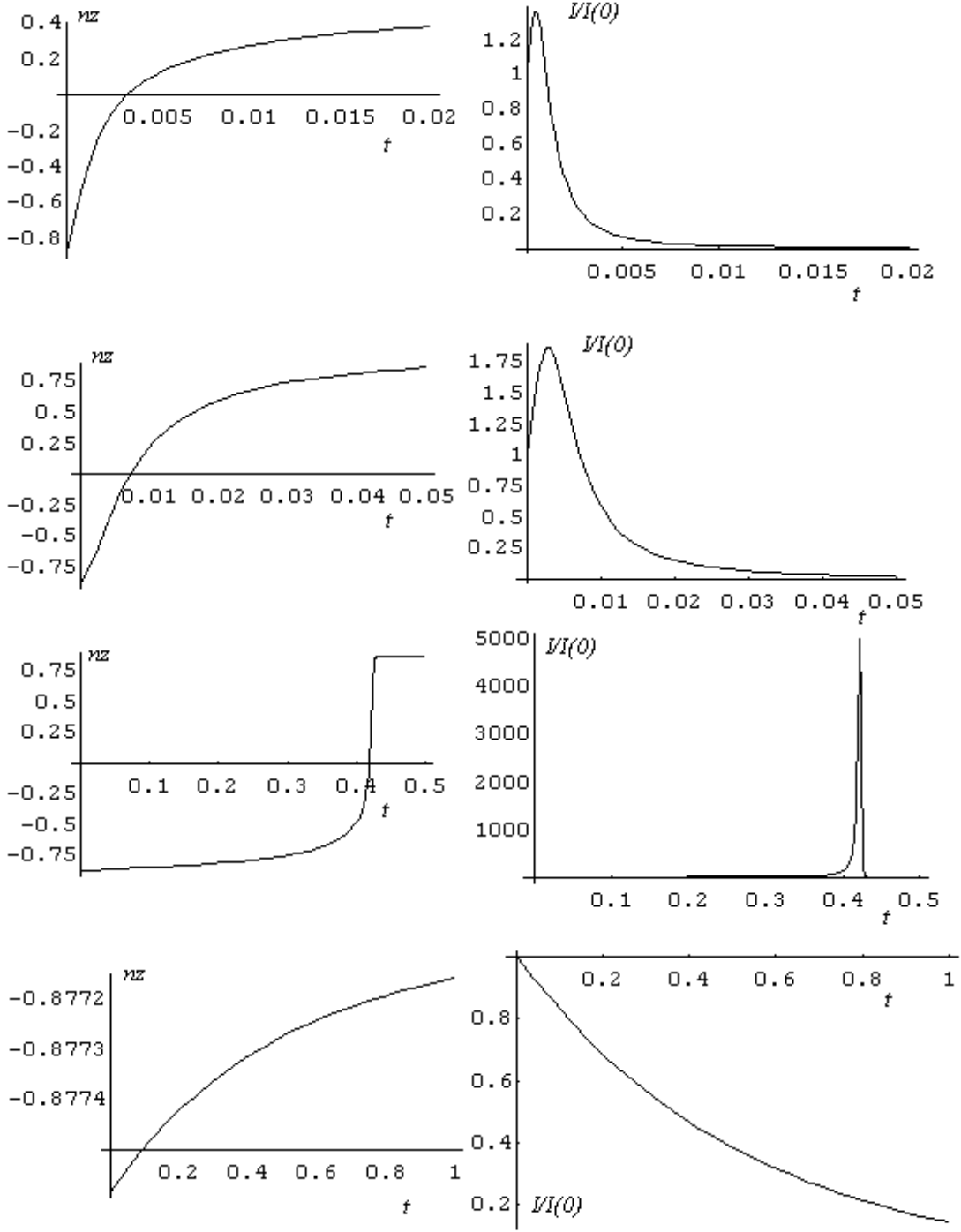}
}
\caption{Evolution of $n_z(t)$ and $I(t)/I(0)$. Parameter
$\xi$ is $100$, the values of parameter $K$ are
$-1.5$, $-0.5$, $0.9$, $1.1$ (up-to-down).}
\label{f2}
\end{figure}

The equations of motion predict a very strong SR pulse
for the value of anisotropy parameter
$K={\beta M(0)\over VH_0}$ close (but smaller) to $1$.
Such a scenario is presented in Fig.\ref{f2}
for $K=0.9$. The delay time is long enough to make the
experimental observation of SR possible.

\newpage
\newpage
\newpage

\newpage
\begin{thebibliography}{99}
\bibitem{l}T. Lis, Acta Crystallogr., Sec. B {\bf 36}, 2042 (1980).
\bibitem{fstz}J.R. Friedman, M.P. Sarachik, J. Tejada, R. Ziolo, Phys. Rev. Lett., {\bf 76}, 3830 (1996).
\bibitem{ppbsc}C. Paulsen, J.-G. Park, B. Barbara, R. Sessoli, A. Caneschi, J. Magn. Magn. Mater. {\bf 140 - 144}, 379 (1995).
\bibitem{hbtcb}J.M. Hernandez, E. del Barco, J. Tejada, E.M. Chudnovsky, G. Bellessa, Europhys. Lett. {\bf 47}, 722 (1999).
\bibitem{d}R.H. Dicke, Phys. Rev. {\bf 93}, 99 (1954).
\bibitem{bbm}P. B\"osiger, E. Brun, and D. Meier, Phys. Rev. Lett., {\bf 38},
602 (1977).
\bibitem{kpsy}Yu.F. Kiselev, A.F. Prudkoglyad, A.S. Shumovskii, and V.I.
Yukalov, Sov. Phys. JETP {\bf 67}, 413 (1988).
\bibitem{bh1} T.S. Belozerova, V.K. Henner and V.I. Yukalov, Phys. Rev.
B {\bf 46}, 682 (1992).
\bibitem{bh2} T.S. Belozerova, V.K. Henner and V.I. Yukalov,
 Comput. Phys. Commun. {\bf 73}, 151 (1992).
\bibitem{cg}E.M. Chudnovsky and D.A. Garanin, Phys. Rev. Lett. {\bf 89},
157201-1 (2002).
\bibitem{ylp}V.I. Yukalov, Laser Phys. {\bf 12}, 1089 (2002).
\bibitem{sk}G.V. Skrotskii and A.A. Kokin, Sov. Phys. JETP {\bf 10}, 572 (1960).
\bibitem{dkh}C.L. Davis, I.V. Kaganov, V.K. Henner, Phys. Rev. B {\bf 62}, 12328 (2000).
\bibitem{nscg}M.A. Novak, R. Sessoli, A. Caneschi, and D. Gatteschi,
J. Magn. Magn. Mater. {\bf 146}, 211 (1995).
\bibitem{gcs}D.A. Garanin, E.M. Chudnovsky, and R. Schilling, Phys. Rev. B
{\bf 61}, 12204 (2000).
\bibitem{ll8}L.D. Landau, E.M. Lifshits,
{\it Electrodynamics of continuous media} ( Pergamon, 1975).
\bibitem{ll2}L.D. Landau, E.M. Lifshits, {\it
The Classical Theory of Field} ( Addison --- Wesley, Cambridge, MS, 1951).
\end{thebibliography}
\end{document}